\documentclass[3p]{elsarticle}
\usepackage{bbm}
\usepackage{graphics}
\usepackage{graphicx}
\usepackage{verbatim}
\usepackage{subcaption}
\usepackage{amsmath, amssymb}
\usepackage{epsfig}
\usepackage[colorlinks]{hyperref}
\usepackage{soul}
\usepackage[utf8]{inputenc}
\usepackage{color}

\usepackage{lineno,hyperref}

\journal{...}

\begin{document}

\begin{frontmatter}

\title{Active matter as the underpinning agency for extraordinary sensitivity of biological membranes to electric fields} 

\author{Anand Mathew}
\author{Yashashree Kulkarni}
\ead{ykulkarni@uh.edu}
\address{Department of Mechanical and Aerospace Engineering, University of Houston, Houston, TX 77204, USA}

\begin{abstract}
  Interaction of electric fields with biological cells is indispensable for many physiological processes. Thermal electrical noise in the cellular environment has long been considered as the minimum threshold for detection of electrical signals by cells. However, there is compelling experimental evidence that the minimum electric field sensed by certain cells and organisms is many orders of magnitude weaker than the thermal electrical noise limit estimated purely under equilibrium considerations. We resolve this discrepancy by proposing a non-equilibrium statistical mechanics model for active electromechanical membranes and hypothesize the role of activity in modulating the minimum electrical field that can be detected by a biological membrane. Active membranes contain proteins that use external energy sources to carry out specific functions and drive the membrane away from equilibrium. The central idea behind our model is that active mechanisms, attributed to different sources, endow the membrane with the ability to sense and respond to electric fields that are deemed undetectable based on equilibrium statistical mechanics. Our model for active membranes is capable of reproducing different experimental data available in the literature by varying the activity. Elucidating how active matter can modulate the sensitivity of cells to electric signals can open avenues for a deeper understanding of physiological and pathological processes.
\end{abstract}


\end{frontmatter}


\section{Introduction}

The study of the interaction of biological cells with electric fields has become increasingly important owing to their ubiquity in physiology, diagnostics, and therapeutics. The role of electric fields spans cell physiology \cite{zhao1999small} and signal transduction \cite{luben1991effects} to targeted drug delivery \cite{ghosh2020active}, tissue engineering \cite{markx2008use}, and  bioelectronic devices \cite{yoon2020nanobiohybrid, huang2024bioelectronics}. 
These electric fields could either be generated by cell processes such as ion transport across the cell membrane, or could be external stimuli and include a wide range of effects depending on the intensity and frequency of the fields. When cells are exposed to sufficient amplitudes the permeability of the membrane is increased \cite{kotnik2019membrane}. This process, known as reversible electroporation,  is temporary, thus allowing cells to reestablish their membrane integrity after some time, and is a valuable tool in biotechnology and medicine, where it facilitates the delivery of otherwise impermeable molecules, such as in electrochemotherapy \cite{campana2019electrochemotherapy}. In contrast, stronger fields (higher pulses of higher amplitude) lead to irreversible electroporation \cite{geboers2020high, napotnik2021cell}, a state where the cell damage is beyond repair and subsequently leads to cell death. Irreversible electroporation has shown promise in cancer treatment \cite{tihomir_georgiev__2023, kai_zhang__2023}, especially for solid tumors such as those in the pancreas, prostate, and liver \cite{bart_geboers_2024}. Broadly speaking, the interactions of electromagnetic fields with biological systems have been of long standing interest due to their role in various phenomena such as animal prey sensing and navigation, embryonic development, and wound repair, medical diagnosis and therapy \cite{Weaver2005}.


A key question that arises in this context is: what is the minimum electric field that a cell can detect and respond to? Despite the extensive exploration of cellular responses to electric fields at high intensities, the minimum electric field a cell or its membrane can detect has relatively remained unexplored experimentally \cite{goodman1983, mcleod1987, cleary1988, lu1994}. Theoretical studies, however, have provided significant insights into this phenomenon. Following ideas from classical signal detection theory, it has long been understood that a cell can detect or respond to an electric signal that exceeds the electrical noise associated with thermal fluctuations and inherent electrical activity within the cellular environment. There have been many pioneering studies on estimating this threshold by calculating the thermal electrical noise for a cell. For instance, Adair \cite{adair1991constraints} estimated this threshold to be 0.02 V for a frequency band of 100 Hz, based on a small section of the membrane with a thickness of 5 nm. Alternatively, another theoretical approach involves considering the membrane as a linear dielectric surface in thermal equilibrium with its surroundings, yielding an estimated noise threshold of 0.36 V for a membrane of the same thickness, with a length of 150 $\mu$m \cite{ahmadpoor2015thermal}. Astonishingly, experimental studies reveal that certain large mammalian cells exhibit extraordinary sensitivity to electric fields and can detect signals that are many orders of magnitude lower than these theoretical predictions for the thermal noise limit \cite{astumian1997}. 

Several efforts have been made to address these inconsistencies. Stochastic resonance has been proposed as a mechanism by which cells may detect signals much lower than the thermal noise limit and has generated significant discussion \cite{astumian1997}. 
More recently, Ahmadpoor et al. \cite{ahmadpoor2015thermal} introduced a model that accounts for the nonlinear dielectric behavior of the membranes. Using variational perturbation methods, they estimated the minimum electrical field threshold detectable by cells. This approach bridges the gap between the theoretical estimates for thermal electrical noise limit and experimental observations, albeit, qualitatively. Although the inclusion of nonlinearity reduces the threshold values dramatically compared to earlier linearized theories, the electric noise limits obtained were still much larger than experimental measurements. 

\begin{figure}[t]
    \centering
    \includegraphics[width=0.7\linewidth]{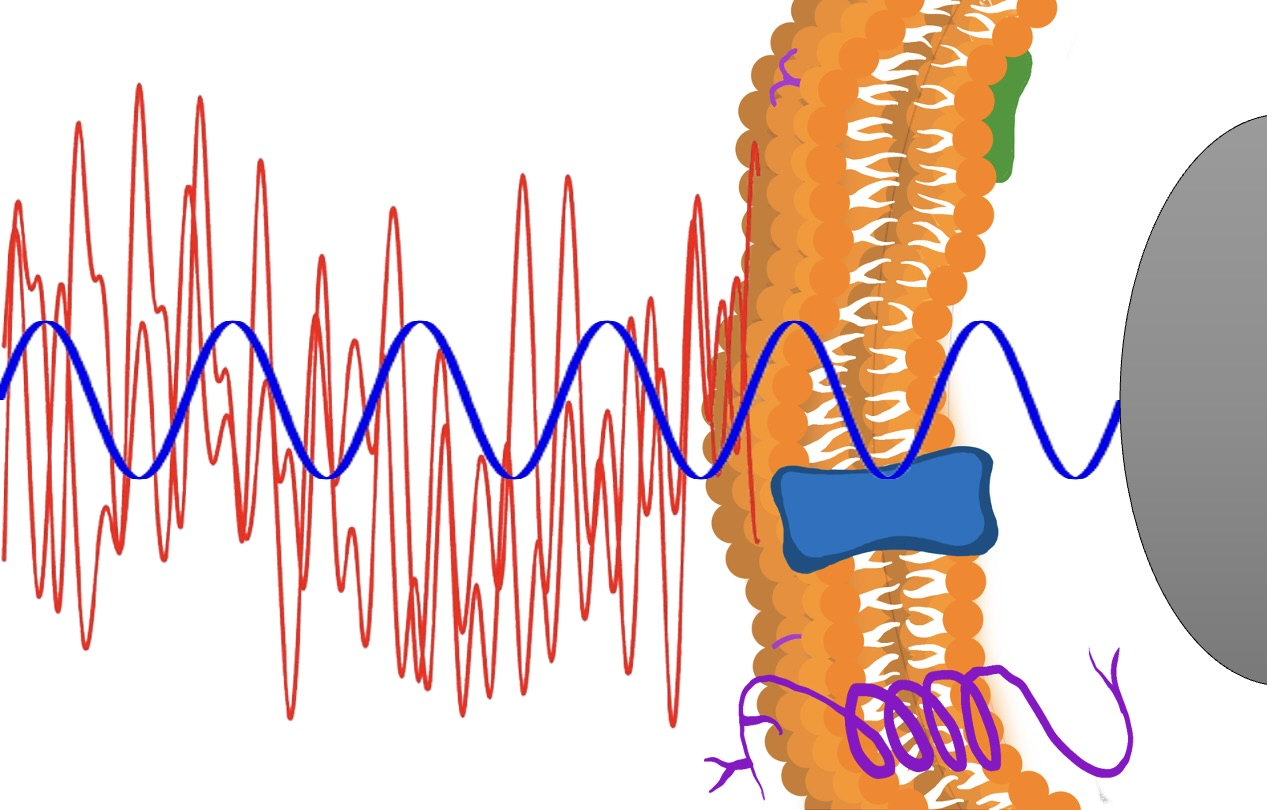}
    \caption{Schematic illustrating that active membranes can detect electrical signals weaker than the thermal electrical noise.}
    \label{fig:schematic}
\end{figure}

In this work, we present the viewpoint that all prior theoretical studies, despite being profoundly insightful, relied on equilibrium considerations. In other words, the membranes considered in these studies exhibit fluctuations solely from thermal vibrations of the membrane molecules, rendering them effectively "passive" or "dead" in a biological sense. 
However, real biological membranes are far more dynamic, incorporating the influence of active agents and processes. Recent work has led to a growing consensus that biological membranes are not merely “passive” surfaces displaying only thermal fluctuations. Instead, they are “active” structures that use internal or external energy sources to perform distinct biophysical functions \cite{rama_2001, rama_2010}. These active membranes incorporate specialized proteins, such as ion channels, that facilitate the transport of molecules such as ions, lipids, or proteins, thereby contributing to various essential physiological processes. As these proteins carry out their roles, they use energy fueled by chemical reactions—such as actin polymerization or ATP hydrolysis—as well as by mechanical stresses, electric fields, or even light \cite{betz_2018} and consequently drive the membrane away from equilibrium. Several recent studies have modeled the role of activity using non-equilibrium statistical mechanics approaches in various physiological processes and phenomena such as active vesicles \cite{girard2005passive, seifert_2012}, vesicle size distribution \cite{kulkarni2023fluctuations, ramesh2024}, active Brownian particles\cite{lee_vella_2017}, active filaments \cite{purohit_active_2020}, and active membranes in electrolytes \cite{lacoste_2006}. 

Taking inspiration from these studies, we develop a theoretical model based on non-equilibrium statistical mechanics framework to examine the role of activity in modulating the sensitivity of active membranes to electric fields. We hypothesize that cell membranes can invoke energy-driven active mechanisms to detect electric signals that are weaker than the theoretical thermal electrical noise limit. To model these driven mechanisms, we move beyond the limitation of equilibrium statistical mechanics and propose a non-equilibrium statistical mechanics approach for the time evolution of polarization in the active membrane in a dynamic environment. This approach enables us to calculate the fluctuations in polarization driven by active components and hence examine the effect of activity on the minimum electric field detected by active membranes. Our hypothesis is consistent with prior experimental studies which have speculated that active mechanisms such as ion channels could play a critical role in imparting extraordinary sensitivity to receptor cells in certain fishes and large mammals \cite{lu1994, astumian1997}. To the best of our knowledge, our work is the first to provide a analytical model to quantitatively demonstrate the role of activity in detection of electric fields by biological membranes and support these experimental observations. 
 
The key contributions of this work are summarized below:
\begin{itemize}
\item We present a theory for active electromechanical membranes based on non-equilibrium statistical mechanics by proposing an over-damped Langevin equation for the time evolution of membrane polarization, analogous to the well-established evolution equation for the out-of-plane deformation of the membrane. 

\item We demonstrate that the dynamic analysis based on the over-damped Langevin equation for polarization in a passive membrane in equilibrium yields the same results for steady-state fluctuation spectra as the conventional equilibrium statistical mechanics approach, say using equipartition of energy. 

\item We elucidate that activity, incorporated into the Langevin equation through the Hamiltonian and as noise, can endow the membrane with the ability to modulate its sensitivity to electric fields as needed and in particular sense electric signals that are deemed undetectable based purely under equilibrium considerations. 

\item We present numerical results to illustrate that the theory is capable of capturing experimental measurements for electric fields sensed by various cell types which are far weaker than the theoretical thermal noise limit.

\end{itemize}

\section{Fluctuations of electromechanical membranes} \label{statmech-polarization}

In order to estimate the minimum electric field detected by a membrane, we need to determine the fluctuations in the electrical field in the membrane as a representative value for the electrical noise in the system. We consider the biological membrane as a linear dielectric and employ tools from statistical mechanics to calculate the fluctuations of polarization in the membrane which can then be used to derive the fluctuations in the electric field.

The study of electromechanical coupling in biological context has garnered significant interest \cite{liping_2013, purohit_flexo_2013, ahmadpoor2015thermal, grassinger2021, torbati_rmp_2022, Pratik2024}. However, these studies relied on employing equilibrium statistical mechanics to study the fluctuations of polarization in membranes. Since active proteins drive a membrane away from equilibrium, accounting for activity requires a dynamic analysis of the fluctuating membrane. In section \ref{passive}, we first present all the elements of the dynamic theory in the case of passive membranes in equilibrium, and then extend it to active membranes in section \ref{active}. We note that in our formulation, the equations for polarization and the mechanical deformation are uncoupled since we do not consider flexoelectricity. Hence, we do not present the theory for the out-of-plane deformation of the membrane which is well-established \cite{rama_2001, lin2006nonequilibrium, betz_2018, kulkarni2023fluctuations, arroyo2009} but focus only on the temporal evolution of the polarization.
 
\subsection{Fluctuations in polarization for passive membranes}
\label{passive}

Consider a dielectric membrane defined over the domain $\mathbb{S} = (0, L)^2$ with thickness $d$ $(d \ll L)$. The membrane is characterized by the state variables: $(P(\mathbf{r}), h(\mathbf{r}))$, where $P(\mathbf{r})$ represents the out-of-plane polarization area density, and $h(\mathbf{r})$ denotes the out-of-plane displacement of the mid-plane of the membrane specified by $\mathbf{r}=(x,y)$. We neglect the in-plane components of the polarization and focus solely on the polarization in the direction perpendicular to the membrane for the sake of simplicity. We postulate the Hamiltonian describing the elastic and electrical energies of the membrane to be
\begin{equation}
\label{eqn:hamiltonian}
    \mathcal{H}^p[P,h] = \int_{\mathbb{S}} \Bigg( \frac{1}{2} \kappa_b (\Delta h(\mathbf{r}))^2 + \frac{1}{2} \sigma(\nabla h(\mathbf{r}))^2 + \frac{1}{2} a |P(\mathbf{r})|^2  \Bigg) \, ds\,.
\end{equation}
Here, $\kappa_b$ is the bending modulus of the membrane. Thus, the first term describes the well-known Canham-Helfrich-Evans bending energy for lipid bilayers \cite{Canham, Helfrich_1973, evans1974bending}. The second term represents the energy penalty for areal changes due to deformation, and $\sigma$ is the associated Lagrange multiplier, also known as the surface tension \cite{safran, Rob, steigmann-book}. The third term accounts for the electrostatic contribution with $a=\left(\frac{1}{\epsilon-\epsilon_0} \right)\frac{1}{d}$, where $\epsilon$ is the permittivity of the linear dielectric membrane. For a rigorous formulation, we should have included the energy associated with the electric field induced by the polarization. However, as demonstrated by Liu and Sharma \cite{liping_2013}, the contribution of this non-local term can be incorporated within the parameter $a$ in the polarization term for biological membranes with permittivity $\epsilon$. We should also have added a flexoelectric electromechanical coupling term in the Hamiltonian. However, as demonstrated by Ahmadpoor et al \cite{ahmadpoor2015thermal}, the effect of flexoelectricity is negligible in the present context, and thus has not been included.

To derive the over-damped Langevin equation for polarization, we consider the membrane to be in a diffusive regime which is a reasonable assumption for a cellular environment. Then, the time rate of change of polarization is proportional to the driving force $F$, 
\begin{equation}
\label{eqn:langevin}
\frac{\partial P(\mathbf{r},t)}{\partial t} = \chi F \,,
\end{equation}
where $\chi$ is a proportionality constant. 
Polarization in a lipid bilayer arises from two phenomena-- ion distribution and dipole reorientation. When ions accumulate on either side of the membrane, they establish an electric double layer, contributing significantly to the membrane's polarization \cite{cevc1983properties}. Additionally, since approximately 80\% of the membrane lipids are zwitterionic, characterized by balanced positive and negative charges, these lipids tend to align with the electric field direction, further polarizing the membrane \cite{mosgaard2014electrical}. Thus, one contribution to the driving force F is the electric field across the membrane which is expressed as the variational derivative of the Hamiltonian of the system with respect to polarization. 

In Brownian dynamics, another contribution to the driving force is a fluctuating force or noise. In the context of polarization, this noise arises from the fluctuations in the electric potential across the membrane due to the thermal fluctuations of molecules causing random movement of the dipoles. We denote this as the polarization noise, $\xi^P$. It is reasonable to assume the noise to be uncorrelated in space and time. Mathematically, we have 
\begin{equation}\label{eqn:polnoise_col}
\langle \xi^{P}(t) \xi^{P}(t') \rangle = B_P \delta(\mathbf{r} - \mathbf{r'}) \delta(t - t')\,,
\end{equation}
where $B_P$ is the strength of the fluctuating electric potential and we will determine it using the fluctuation-dissipation theorem. We can now write the over-damped Langevin equation for membrane polarization as
\begin{equation}
\frac{\partial P(t)}{\partial t} = \chi \left(- \frac{\delta\mathcal{H}[P, h] }{\delta P} + \xi^P(t)\right).
\end{equation}s
The proportionality constant, $\chi$, is a function of the electrical properties of the membrane. From the RC circuit equivalence of the membranes, we propose that the value of $\chi$ depends on the resistance ($R$), capacitance ($C$), permittivity ($\epsilon_0$), and thickness of the membrane. Thus, $\chi=\frac{1}{a R C}$ is a dimensionally accurate value for the constant. It is convenient to work with the Fourier transform of the equation which is obtained as
\begin{equation}
\label{eqn:langevin_pol_fourier}
    \frac{\partial P_\mathbf{q} (t)}{\partial t} = \chi_{\mathbf{q}} \left[-a P_\mathbf{q} (t)  + \xi_\mathbf{q}^{P}(t) \right] \,,
\end{equation}
where 
\begin{equation}
P_\mathbf{q} = \frac{1}{L} \int_{\mathbb{S}} d\mathbf{r} \, P(\mathbf{r}) e^{-i \mathbf{q} \cdot \mathbf{r}} \quad \text{and} \quad
P(\mathbf{r}) = \frac{1}{L} \sum_\mathbf{q} P_\mathbf{q} e^{i \mathbf{q} \cdot \mathbf{r}}.
\end{equation}
Solving equation \eqref{eqn:langevin_pol_fourier} and taking the limit $t\rightarrow\infty$, we avoid artifacts due to initial condition and evaluate the mean-square fluctuations for polarization as \cite{zwanzig2001}:
\begin{equation}
\label{eqn:Pq_int}
    \langle |P_{\mathbf{q}}|^2 \rangle = \chi_{\mathbf{q}}^2 \lim_{t \to \infty} \int_0^t ds_1 \int_0^t ds_2 \langle \xi_\mathbf{q}^{P}(s_1) \xi_\mathbf{q'}^{P}(s_2) \rangle e^{\omega^{P}_{\mathbf{q}}(s_1 + s_2 - 2t)}
\end{equation}
Here, $\omega^{P}_\mathbf{q}=\chi_{\mathbf{q}} a$. Substituting the correlation \eqref{eqn:polnoise_col} in equation \eqref{eqn:Pq_int} we get,
\begin{equation}
    \langle |P_{\mathbf{q}}|^2 \rangle = \chi_{\mathbf{q}}^2 \frac{B_{P}}{\omega^{P}_{\mathbf{q}}}.
\end{equation}
According to the fluctuation-dissipation theorem, at $t\rightarrow\infty$, the autocorrelation function for polarization must be equal to that derived from equilibrium statistical mechanics using equipartition of energy leading to the relation $B_{P} = k_B T/L^2 \chi_{\mathbf{q}}$. Thus, the fluctuation spectrum for polarization of a passive membrane is obtained as
\begin{equation} \label{eq:passive}
    \langle |P_{\mathbf{q}}|^2 \rangle = \frac{k_B T}{L^2 a}.
\end{equation}
which is identical to the expression obtained from equilibrium statistical mechanics approach \cite{liping_2013} as ensured by the fluctuation-dissipation theorem \cite{zwanzig2001}. We also observe that the autocorrelation for polarization for a passive membrane in steady state depends neither on the wavevector nor the coefficient $\chi$. 

\subsection{Fluctuations in polarization for active membranes}
\label{active}
Active membranes differ fundamentally from passive membranes due to their dynamic and interactive nature, driven by the activity of active agents embedded within the membrane. While passive membranes respond to external stimuli through basic elastic deformations and electrostatic effects, active membranes exhibit additional complexity arising from molecular processes and energy-consuming mechanisms. 
To extend the Langevin equation derived in the previous section for active electromechanical membranes, we hypothesize that the effect of active mechanisms can be modeled through two distinct contributions.   

First, we note that active membranes are highly heterogeneous and densely populated with proteins, many of these being active proteins that can exert their influence through long-range interactions. Ion pumps and channels, which are known to be active proteins \cite{betz_2018}, engage in transporting ions across the membrane, actively maintaining concentration gradients \cite{yuri_gaididei_1993,david_l_armstrong_1998} and causing membrane conformational changes through interaction with other proteins \cite{tillman2003effects}. These ion movement and protein-protein interactions further lead to changes in membrane polarization. We incorporate this non-local effect by adding a contribution to the Hamiltonian which is proportional to the spatial gradient of the polarization. Thus, the modified Hamiltonian, $H^a$, for active membranes becomes
\begin{align}
\label{eqn:updated_hamiltonian}
\mathcal{H}^{a}[P,h] = \int_{\mathbb{S}} \Bigg( \frac{1}{2} \kappa_b (\Delta h(\mathbf{r}))^2 + \frac{1}{2} \sigma(\nabla h(\mathbf{r}))^2 + \frac{1}{2} a |P(\mathbf{r})|^2 \notag \\
+ \frac{1}{2} \beta |\nabla P(\mathbf{r})|^2 \Bigg) \, ds
\end{align}
Second, under the influence of a fluctuating external electric field, proteins can undergo structural rearrangements that are often linked to the proteins' electric dipoles, which reorient in response to the external field \cite{bezanilla2008membrane}. Additionally, integral and peripheral proteins are asymmetrically distributed and oriented within the membrane \cite{rothman1977membrane}. These proteins carry positive and negative charges, creating an electric dipole that fluctuates with the external electric field.  As a result, the interplay between protein conformational changes, electric dipole fluctuations, and ion transport mechanisms contributes to the dynamic regulation of the potential across the membrane \cite{mosgaard2014electrical}. All these effects are captured in the Langevin equation by adding an active noise term, $\eta_\mathbf{q}^{P}$. The active polarization noise is considered to be uncorrelated in space and exponentially correlated in time. This is a reasonable assumption inspired by prior theoretical studies on the mechanical response of active membranes \cite{lin2006nonequilibrium, seifert_2012, kulkarni2023fluctuations}. Thus, the active polarization force is assumed to have zero mean and the following form for the autocorrelation function:
\begin{equation}
\langle \eta_\mathbf{q}^{P}(t) \eta_\mathbf{q'}^{P}(t') \rangle =\Gamma_\mathbf{q}^{P} \delta(\mathbf{q}-\mathbf{q'}) \exp{\left(-\frac{|t-t'|}{\tau^{P}}\right)}
\end{equation}
where $\tau^{P}$ is the correlation time associated with the active process. The constant $\Gamma_\mathbf{q}^{P}$ is the amplitude of active noise or in this case the squared of the electric field generated by the active components of the membrane.

Using the modified Hamiltonian from  \eqref{eqn:updated_hamiltonian} we derive the Langevin equation for an active membrane in Fourier space as
\begin{equation}
\label{eqn:act_lang_pol}
\frac{\partial P_\mathbf{q} (t)}{\partial t} =  - \omega_\mathbf{q} P_\mathbf{q} (t) + \chi_{\mathbf{q}}\xi_\mathbf{q}^{P}(t) + \chi_{\mathbf{q}}\eta_\mathbf{q}^{P}(t) \,,
\end{equation}
where, $\omega_\mathbf{q} = \chi_{\mathbf{q}} \left(a +  \beta q^2\right)$. 
Solving equation \eqref{eqn:act_lang_pol} and then evaluating the mean-square fluctuations at time $t\rightarrow\infty$ we arrive at,
\begin{align}
    \langle |P_{\mathbf{q}}|^2 \rangle = \chi_{\mathbf{q}}^2 \lim_{t \to \infty} \int_0^t ds_1 & \int_0^t ds_2 \left(\langle \xi_\mathbf{q}^{P}(s_1) \xi_\mathbf{q'}^{P}(s_2) \rangle + \right.  \nonumber\\
    & \left. \langle \eta_\mathbf{q}^{P}(t) \eta_\mathbf{q'}^{P}(t') \rangle\right) e^{\omega_{\mathbf{q}}(s_1 + s_2 - 2t)} \label{eqn:act_Pq_int}
\end{align}

Substituting the noise correlation and evaluating the integral we get the fluctuations in polarization in an active membrane,
\begin{equation}\label{eq:active}
\langle| P_\mathbf{q}|^2\rangle = \frac{k_B T}{L^2  \left(a +  \beta q^2\right)} + \frac{\Gamma_\mathbf{q}^{P} \chi_{\mathbf{q}}^2 }{\omega_{\mathbf{q}} \left(\omega_{\mathbf{q}} + \frac{1}{\tau^{P}}\right)}.
\end{equation}

This is a key result of our paper. First, it reveals that active noise enhances the polarization fluctuations. This is consistent with prior observations that active noise increases mechanical fluctuations \cite{lin2006nonequilibrium, kulkarni2023fluctuations}. Second, it shows that active mechanisms that contribute through the spatial gradient of polarization decrease the polarization fluctuations. In subsequent sections, we will find that it is the polarization gradient term which plays a role in determining the minimum threshold for detection of electric fields. 

\section{Fluctuations in electric field} \label{electrical-noise}

Following the work of Ahmadpoor et al. \cite{ahmadpoor2015thermal}, we can use the results from the previous section to estimate the root mean square of the electric field and the frequency-dependent electrical noise. The latter is needed to compare our results with measurements since experiments indicate the noise threshold to be sensitive to the frequency of the applied field. Here, for the sake of convenience, we summarize the general approach and the final expression presented by Ahmadpoor et al. \cite{ahmadpoor2015thermal}  and refer the reader to their paper for details. 

To calculate the electric field generated by a polarized membrane, we apply the Maxwell's equation under non-conducting boundary conditions as
\begin{equation}
\text{div}\left( -\epsilon_0 \nabla \phi + \frac{P}{d} \nu(z) \mathbf{e}_z \right) = 0,
\end{equation}
where,
\begin{equation}
\nu(z) = 
\begin{cases} 
1 & \text{if } z \in \left[ -\frac{d}{2}, \frac{d}{2} \right] \\
0 & \text{otherwise}\,,
\end{cases}
\end{equation}
$\phi$ is the potential across the membrane and $d$ is the membrane thickness. The solution can be conveniently found in Fourier space. As remarked by Ahmadpoor et al., the final results for fluctuations of electric field are insensitive to the boundary conditions. Under conducting boundary conditions, the autocorrelation function for the potential is obtained as
\begin{equation} \label{eq:electrical-noise}
\langle V^2 \rangle = \frac{1}{\epsilon_0^2} \int_{q_{min}}^{q_{max}} \langle| P_\mathbf{q}|^2\rangle.
\end{equation}

\begin{figure}[t]
    \centering
    \includegraphics[width=0.8\linewidth]{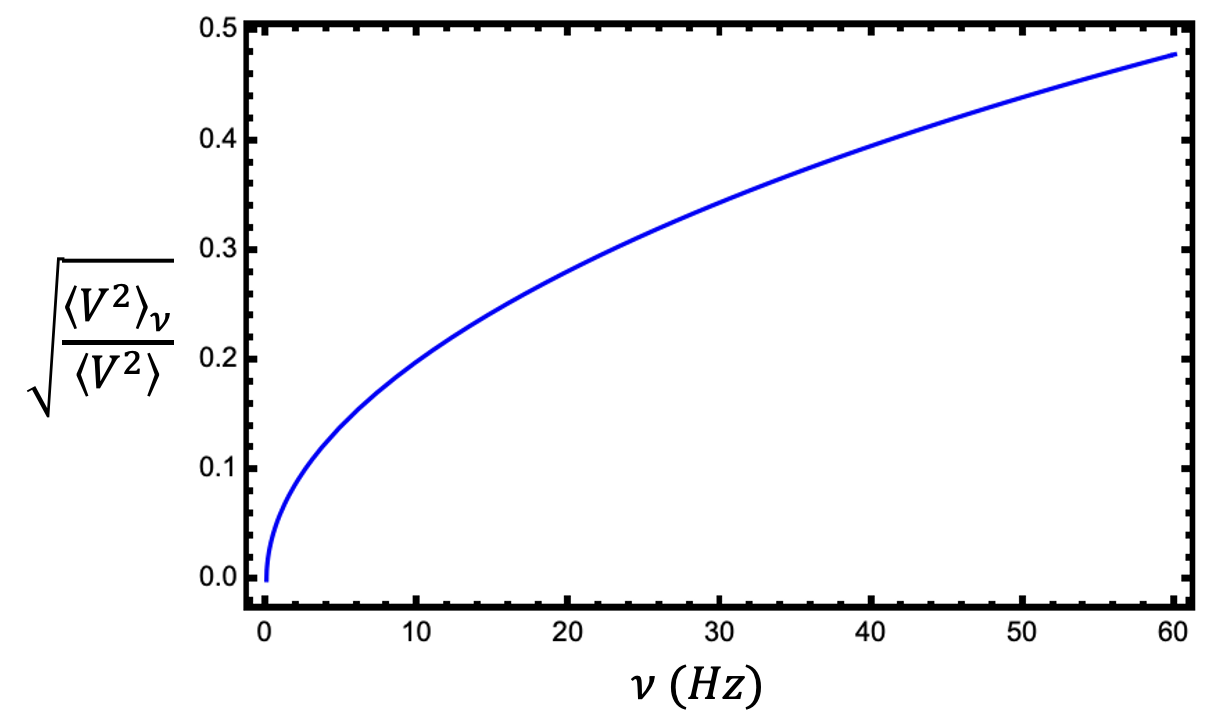}
    \caption{Calculated thermal noise in the presence of time-dependent fields. The relaxation time is considered to be 1 ms.}
    \label{fig:freqplot}
\end{figure}

\eqref{eq:electrical-noise} is a significant result by Ahmadpoor et al. \cite{ahmadpoor2015thermal} as it provides a way to express electrical noise in terms of the fluctuation spectra for polarization of the membrane. Using \eqref{eq:passive} for passive membrane or \eqref{eq:active} for active membranes gives the threshold electrical field in either case. 

Although \eqref{eq:electrical-noise} suffices to estimate the electrical noise limit theoretically, we need frequency dependent values for the electric field fluctuations to draw comparisons with experimental data. Again, following Ahmadpoor et al. \cite{ahmadpoor2015thermal}, we calculate the power spectrum of the fluctuating electric field as
\begin{equation}
G(\nu) = 4 \int_{0}^{\infty} \langle V^2 \rangle e^{-t/\tau} \cos(2\pi \nu t) \, dt = \frac{4 \langle V^2 \rangle \tau}{1 + (2\pi \nu \tau)^2}
\end{equation}
where $\nu$ is frequency and $\tau$ is the relaxation time. $G(\nu)$ is equivalent to the Nyquist noise for a resister and is obtained by integrating it over the frequency. 
Let us assume that the frequency we are interested in is $\nu_1$. Then, integrating the above equation from 0 to $\nu_1$ yields the frequency-dependent fluctuation spectra for the electric field as
\begin{equation}\label{eq:freq-electric-noise}
    \langle V^2 \rangle_\nu = \frac{2 \langle V^2 \rangle}{\pi}\tan^{-1}(\frac{\pi \nu_1}{500})
\end{equation}
Here, the relaxation time for biological membranes is assumed to be 1 millisecond \cite{ahmadpoor2015thermal}. Figure \ref{fig:freqplot} shows the variation of the root mean square of the electrical noise obtained in \eqref{eq:freq-electric-noise} with frequency. It is worth emphasizing that at very low frequencies in the range 0 to 10 Hz, the noise threshold $\langle V^2 \rangle_\nu$ can be much smaller than $\langle V^2 \rangle$.

\section{Numerical results and experimental validation}
\label{results}

To understand the effect of the two different active mechanisms considered in this study, we present numerical results for the variation of $\langle V^2 \rangle$ with $\beta$ and $\Gamma^P_q$. 
For the purpose of the numerical calculations, we assume the membrane length to be $L = 150$ nm and membrane thickness as $d = 5$ nm. Temperature is taken to be 300 K, and the membrane permittivity is taken as $\epsilon = 2\epsilon_0$. For the activity noise term, the relaxation time for the active process is taken to be $\tau_P = 5$ ms and the coefficient $\chi$ is estimated as $\chi = \frac{1}{aRC}$ with $R = 2.12 \times 10^6 \Omega \cdot m $ and $C = 0.54 \times 10^{-6}\, F/m $ for the membrane. 

\begin{figure}[t]
    \centering
    \includegraphics[width=0.8\linewidth]{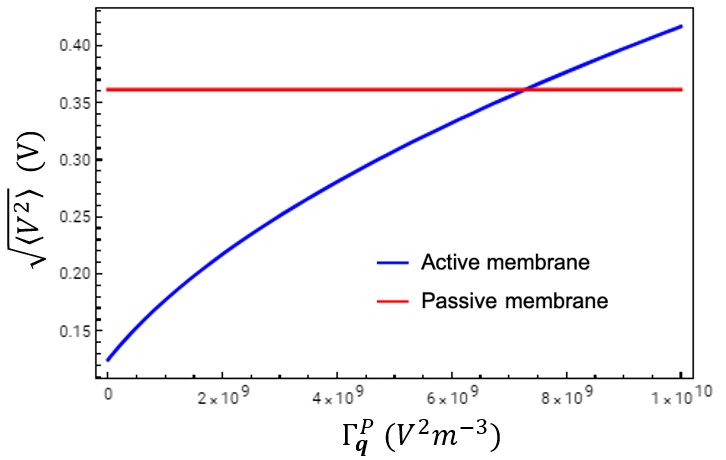}
    \caption{Electrical noise for active membranes as a function of the active noise strength ($\Gamma_\mathbf{q}^{P}$) shown by the blue curve. The value of noise of the passive membrane is 0.36V (red curve). $\beta$ is set to $a\delta^2 = 2252.82 F^{-1}m^{2}$ with $\delta = 10 nm$.}
    \label{fig:gamma_vary_plot}
\end{figure}

\begin{figure}[t]
    \centering
    \includegraphics[width=0.8\linewidth]{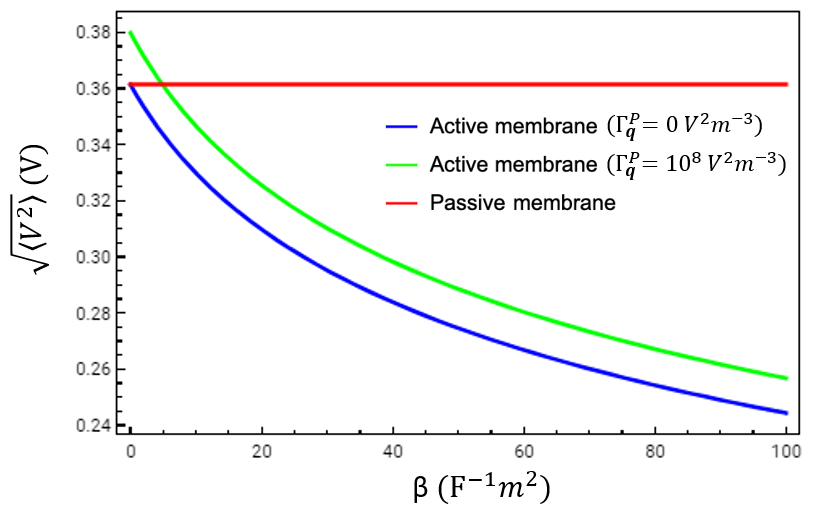}
    \caption{Electrical noise for active membranes as a function of $\beta$ with (green curve) and without (blue curve) active noise. The value of noise of the passive membrane is 0.36V (red curve).}
    \label{fig:beta_vary_plot}
\end{figure}

Figure \ref{fig:gamma_vary_plot} shows that increase in active noise increases the threshold electrical noise. Here, the value of $\beta$ is held fixed and chosen to be on the order of  $a\delta^2$ where $\delta$ is regarded as the inter-protein distance for the passive case and estimated to be around 10 nm \cite{ahmadpoor2015thermal}. Thus, it is clear that active noise is not the key player in the heightened sensitivity of certain biological membranes to very weak electric fields. In contrast, Figure \ref{fig:beta_vary_plot} shows that the threshold electrical noise decreases dramatically with increasing $\beta$. This quantitatively demonstrates that the non-local effect induced by active proteins such as ion channels through the spatial gradient of polarization could be an underpinning mechanism that active membranes employ to reduce the electrical noise and possess extraordinary sensitivity to far weaker electric fields as necessary. 

Finally, we compare our results with different experimental studies from literature. For the sake of illustration, we select two studies by Clearly et al. \cite{cleary1988} and McLeod et al. \cite{mcleod1987} respectively which investigated the sensitivity of different mammalian cells to extremely weak electric fields at different frequencies. Since active noise increases fluctuations, which in turn results in larger thermal electrical noise, we set $\Gamma_q^P$ to be zero. Using \eqref{eq:active} and \eqref{eq:electrical-noise}, we estimate the value of $\beta$ by fitting to the experimental values. As seen in Table \ref{tab:comparison}, the minimum electric field estimated by the linear dielectric model based on equilibrium statistical mechanics (which yields a frequency-independent value of 0.36 V) is orders of magnitude larger than experimental measurements. Although the nonlinear dielectric model \cite{ahmadpoor2015thermal} based on equilibrium thermal fluctuations performs significantly better than the simplistic linear model, the threshold values are still substantially larger (by an order of magnitude) from experimental values. In striking contrast, incorporating active mechanisms in the theory for polarization of membranes enables the present model to approach experimental values for extremely low electric fields sensed by cells simply by varying the activity. 

\section{Conclusion} \label{conc}

In summary, we investigated the role of activity in the sensitivity of biological membranes (and cells) to electric fields by way of a non-equilibrium statistical mechanics based model. Our study reveals that activity, attributed to different sources, can endow the membranes with the ability to sense and respond to electric fields that are are far weaker than the thermal noise limit estimated from purely equilibrium statistical mechanics. Our model for active membranes is capable of reproducing different experimental data available in the literature by varying the activity and thus provides a possible resolution for a long-standing discrepancy between theoretical studies and experimental measurements for the minimum electric field that can be detected by cells. As part of future study, we intend to enrich the model to incorporate electromechanical coupling through flexoelectricity and investigate its role in the interaction of active membranes with electric fields. Understanding electromechanical biological phenomena that may be impacted by increased fluctuations due to active noise in polarization is another interesting direction of study. Elucidating how cells can modulate their interaction with electric signals through activity can proffer deeper insights into physiological processes and open avenues for novel applications in biotechnology and medicine.

\begin{table*}
\centering
\begin{tabular}{|c|c|c|c|c|c|}
\hline
\vtop{\hbox{\strut Frequency} \hbox{\strut {(Hz)}}} & \vtop{\hbox{\strut Experimental} \hbox{\strut values (V/cm)}} & \vtop{\hbox{\strut Linear dielectric} \hbox{\strut values(V/cm)}} & \vtop{\hbox{\strut Nonlinear dielectric} \hbox{\strut model (V/cm) \cite{ahmadpoor2015thermal}}} & \vtop{\hbox{\strut Present} \hbox{\strut model (V/cm)}} & $\beta$ (F$^{-1}$m$^{2}$) \\ \hline
1                       & 0.6 \cite{cleary1988}      & $4.56 \times 10^4$                                                                     & 100                                                                 & 0.6                                                                 & $2.49 \times 10^{15}$       \\ \hline
0.1                     & 300 \cite{mcleod1987}      & $1.44 \times 10^4$                                                                       & 30                                                                  & 300                                                                & $1.64 \times 10^{8}$        \\ \hline
1                       & 2.1 \cite{mcleod1987}      & $4.56 \times 10^4$     &   100                                                                  & 2.1                  &    $2.03 \times 10^{14}$                         \\ \hline
$10^1$                       & 1.5 \cite{mcleod1987}      & $1.44 \times 10^5$                                                                    & 300                                                           & 1.5         &      $3.95 \times 10^{15}$                        \\ \hline
$10^2$                       & 30  \cite{mcleod1987}    & $4.30 \times 10^5$                                                                         &  900                                                                    & 30       &  $8.89 \times 10^{13}$                           \\ \hline
$10^3$                        & 600  \cite{mcleod1987}    & $6.83 \times 10^5$                                                                       &  1400                                                                     & 600     &   $5.56 \times 10^{11}$                           \\ \hline
\end{tabular}
\caption{Comparison of our results with experimental values and values from prior modeling studies available in literature. The last column shows the values of $\beta$ required to obtain the experimental values. $\Gamma_q^P$ is taken to be zero.}
\label{tab:comparison}
\end{table*}

\section*{Acknowledgement} 
The authors gratefully acknowledge the support of NSF under grant CMMI--2227556. We thank Professor Pradeep Sharma for insightful discussions on the electromechanical behavior of membranes.


\bibliography{references}

\end{document}